**Mechanosensitive polymer matrices of biologically-relevant compliance based on upconverting nanoparticles**

*Cindy H. Shi\*, Mia C. Cano, Jason R. Casar, Parivash Moradifar, Beatriz G. Robinson, Julia A. Kaltschmidt, Miriam B. Goodman, Jennifer A. Dionne\**

Cindy H. Shi, Jason R. Casar, Parivash Moradifar, Jennifer A. Dionne
Department of Materials Science and Engineering, Stanford University, Stanford, CA 94305, USA.
Emails: cinshi@stanford.edu, jcasar@stanford.edu, pmoradi@stanford.edu, jdionne@stanford.edu
ORCID: 0000-0001-8461-156X

Mia C. Cano
Department of Physics, Stanford University, Stanford, CA 94305, USA.
Email: miacano@stanford.edu

Beatriz G Robinson
Neurosciences IDP Graduate Program, Stanford University School of Medicine, Stanford, CA, USA.
Email: beatrizr@stanford.edu

Julia A Kaltschmidt
Department of Neurosurgery, Stanford University School of Medicine, Stanford, CA, USA.
Email: jukalts@stanford.edu

Miriam B. Goodman
Dept of Molecular and Cellular Physiology, Stanford University, Stanford, CA 94304 USA
Email: mbgoodmn@gmail.com
ORCID: 0000-0002-5810-1272

The authors have no conflicts of interest in the findings and publishing of this article.

Data supporting the findings of this study were generated at Stanford University, and are available from the corresponding author (C.H.S.) upon request.

This work was supported by the Chan-Zuckerberg Initiative, Q-NEXT (U.S. Department of Energy Office of Science National Quantum Information Science Research Centers under Award No. DE-FOA-0002253), National Science Foundation Graduate Research Fellowships Program (NSF), and the Department of Defense National Defense Science and Engineering Graduate Fellowship (NDSEG).

**Abstract**

Upconverting nanoparticles (UCNPs) are promising optical biomechanical force sensors due to their near-infrared excitation, low toxicity, photostability, and linear colorimetric sensitivity to microNewtons of force. Recently, a composite force sensor based on UCNPs embedded in a polystyrene microbead enabled the first real-time measurement of feeding forces in living nematodes. However, the comparatively large stiffness of polystyrene only makes it relevant to biomedical application in a small subset of biological tissue. To facilitate deployment of UCNPs into biological tissues with a range of mechanical properties, we expand upon polymer-UCNP composite systems by embedding UCNPs in three polymer matrices with varying stiffnesses (epoxy resin, polydimethylsiloxane, and alginate hydrogels). Furthermore, to enhance these composites' mechanosensitivity, we methodically investigate using two different core-shell architectures of SrLuF-based UCNPs doped with ytterbium, erbium, and varying manganese concentrations. We calibrate polymer-UCNP composite optical force sensitivity with colocalized atomic force and confocal microscopy. Using the red to green emission ratio (Δ% $I_{Red}$:$I_{Green}$) as the force read-out, we determine that SrLuF:Yb$_{0.28}$Er$_{0.025}$Mn$_{0.013}$ @ SrYF dispersed in epoxy resin exhibits the greatest emission color change (12 Δ%$I_{Red}$:$I_{Green}$ per microNewton). Finally, we map forces in the epoxy-UCNP composite on the macroscale between the joint of a chicken wing bone using a commercially available wide field microscope, thereby demonstrating its ability to optically measure pressures *in situ*. This work establishes the utility and modularity of the UCNP-polymer composite system for force-sensing in geometrically and mechanically diverse biological systems.

1. **Introduction and Background**

Mechanical forces and pressures are foundational to how biological tissues function. Whether tissues are stiff like bones (elastic modulus of ~GPa) or ligaments (elastic modulus ~MPa), soft like internal organs (elastic moduli ~5-20 kPa), or even softer like brain tissue and skin (elastic moduli ~1 kPa), their structural integrity is sensitive to health status, age, and pathophysiology[1,2]. It is well-established that mechanical stress aids in improving bone density and stimulates bone formation, while lack of physical strain leads to fragile bones and disuse osteoporosis[3]. In the gastrointestinal tract, the magnitude and rhythm of bowel wall deformation dictates proper digestion, absorption, and propulsion; abnormal increases in intestine luminal pressure or interstitial pressure between cells are markers of gut motility disorders such as irritable bowel syndrome or Crohn's disease[4]. Another example of mechanical pathophysiological change is that normally viscoelastic brain tissue stiffens during cancer progression due to cell extracellular matrix composition changes, cell compression, and elevated fluid pressure[5]. Mapping these forces *in situ* and in real time, can help uncover the role of mechanotransduction in disease phenotypes at the tissue level, enhancing treatments and informing fundamental research directions.

Quantification of forces within or produced by biological tissues requires sensors that are deployable *in situ* and perturb the native biological environment as little as possible[6]. Current mechanical testing methods on biological tissues are usually *ex vivo* methods. Atomic force microscopy (AFM) can elucidate information on biomolecular mechanics; however, it is a surface technique that cannot access internal forces non-invasively, and current tip-to-cell contact models can oversimplify cell mechanics[6-8]. Nanoindentation, a technique similar to AFM, is only accurate for stiffer, calcified tissues[6]. *In vivo* techniques often have limited generalizability and resolution as well. Wearable ultrasonic transducer arrays can calculate tissue modulus from wave velocity but depend on assumptions of tissue density, and transducer size limits spatial resolution[1,2]. Magnetic resonance elastography can achieve deep tissue mechanical data with one millimeter resolution, but requires a high-cost elaborate setup only available in hospitals[1,2]. Catheter-based implantable piezoelectric force sensors are also a direct method of measuring forces commonly applied to cardiovascular and urethral systems, but require extremely precise surgical manipulation of the long, thin catheter to probe forces without damaging tissue[9,10]. Their fabrication is also complex, delicate, and costly[9]. Thus, there is a need for more accessible, generalizable, remotely read biological mechanosensors with greater spatial resolution, potentially in the form of smart implants or miniaturized wearables.

Upconverting nanoparticles (UCNPs) are especially well-suited candidates for optical mechanosensing in biological systems. Unlike conventional downconverting fluorophores, UCNPs utilize near-infrared excitation, a wavelength range where tissue absorbance and scattering are minimized. This enables enhanced penetration depths and reduced autofluorescence. Furthermore, compared to organic fluorophores or molecular dyes for sensing biomechanical forces, such as zinc porphyrins or FRET-based molecular tethers, UCNPs show stable optical signal and no photobleaching even over hours of irradiation[7,8]. UCNPs have significantly better biocompatibility than heavy metal-containing quantum dots[7,8]. Indeed, they elicit no measurable toxicity in a *C. elegans* model when delivered alone or encapsulated in polymers[11-13], and previous studies have found no immediate cytotoxicity of surface-modified UCNPs to mammalian fibroblast and kidney cell lines[14–16]. Recently, we developed a UCNP-based polystyrene mechanosensor that rapidly reports stress changes in the *C. elegans* feeding organ[13]. However, polystyrene is stiffer than many biological tissues. To realize UCNP-based force sensing in a wider range of biological tissues, mechanosensors must incorporate more compliant polymer hosts to use them in natural tissues. Here, we show that our UCNP-polymer platform retains mechanosensitivity across a range of biologically relevant polymer matrix stiffnesses, opening the door to a range of future mechanosensitive implants with tunable compressibility.

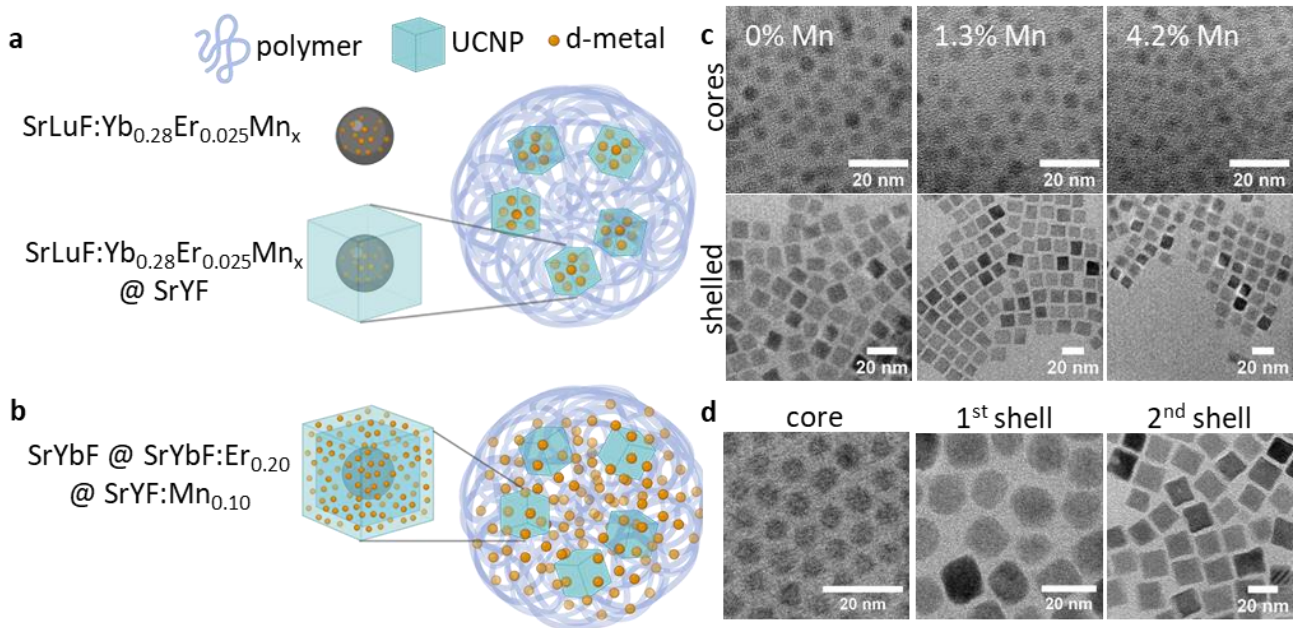

Figure 1. Illustration of two schemes used to incorporate UCNPs and Manganese ions in a polymer matrix and synthesized UCNP electron micrographs. (a) Mn ions are doped into the UCNP core lattice. (b) Mn ions present in the outer UCNP shell and as cross-linkers within the polymer matrix. (c) Transmission Electron Micrograph (TEM) images of UCNP cores and shelled of various Mn dopant concentrations. (d) TEM images of UCNP core and shell layers designed for transfer to external Mn ions

Previous work on pressure-sensitive UCNPs has shown that nanoparticles doped with the Yb-Er sensitizer-emitter pair, emit sharp peaks in the red (660 nm) and green (525nm and 545nm) visible regimes when irradiated with 980 nm light. The ratio of the integrated intensities of these two peaks ($I_{Red}:I_{Green}$) changes linearly and without hysteresis as a function of externally applied pressure[17,18]. This change is primarily due to increased Er-Er and Er-Yb cross relaxation rates as the ceramic lattice is deformed interionic distances change, quenching the green-emitting states more effectively than red-emitting states and causing a colorimetric change in UCNP emission[13]. Doping a d-metal into the octahedrally coordinated sites of the cubic lattice can further sensitize energy transfer from the green-emitting to red-emitting states to pressure changes[19]. This doping thus enhances the UCNP colorimetric pressure sensitivity[19]. Previous tests on a series of alternate host lattices found that the SrLuF host lattice resolves the smallest changes in hydrostatic pressure when compressed in a diamond anvil cell[18]. Thus, UCNP dopant concentration and host lattice types can tune pressure sensitivity, but this has been demonstrated only with high pressures (~GPa) in a diamond anvil cell. Recent work showed that UCNPs embedded in biocompatible polymer retain their mechanosensitivity at lower forces and pressures (~microNewtons and ~MPa) while providing a geometrical format more suitable for the targeted biological system[13].

Building upon these previous studies, we aim to create the most force-sensitive SrLuF host lattice UCNP-polymer composites for various biological tissue compliances. We approach this goal in three stages: fabrication, characterization, and demonstration. First, we optimize composite mechanosensitivity by modifying UCNP architecture and Mn localization in polymer host matrices of variable stiffnesses. To investigate the effect of Mn doping within the UCNP cores on UCNP force sensitivity, we synthesized $SrLu_{0.695-x}F:Yb_{0.28}Er_{0.025}Mn_x$ @ SrYF UCNPs. We then embedded these UCNPs in two polymer matrices: 1) the softer polydimethylsiloxane (PDMS) and 2) the stiffer epoxy resin . To investigate the effect of spatially compartmentalized Mn ion presence on UCNP force sensitivity, we synthesized SrYbF @ $SrYbF:Er_{0.20}$ @ $SrYF:Mn_{0.10}$ UCNPs to maximize energy migration between Er emitters and external Mn ions, then embedded these UCNPs into alginate hydrogels crosslinked with either Mn or Ca divalent cations . For characterizing force sensitivity of all these UCNP-polymer composites, we use simultaneous atomic force and confocal microscopy (rather than a diamond anvil cell) to achieve higher stress resolution at more biologically relevant magnitudes. We vary either the internal Mn dopant concentration or the external polymer matrix crosslinking ion type to achieve the greatest $\Delta\%I_{Red}:I_{Green}$. Finally, we demonstrate the utility of this technology for spatial force mapping on a wide field microscope under two conditions, beneath weighted beads and between the bones of a chicken wing.

## 2. Results and Discussion

In designing force-dependent UCNP-polymer composites, we draw on previous observations from literature that Er-Mn interionic interaction is important for modulating energy transfer behavior and therefore emission color[19–23]. $Mn^{2+}$ has a broad $^4T_1 \rightarrow {^6A_1}$ emission band that overlaps with both the narrow green ($^2H_{11/2} + {^4S_{3/2}} \rightarrow {^4I_{15/2}}$) and red ($^4F_{9/2} \rightarrow {^4I_{15/2}}$) emissions of $Er^{3+}$, enabling successive depopulation of the green-emitting state and population of the red-emitting state[22]. Increasing pressure would feasibly reduce the Er-Mn interionic separation, and broaden the Mn emission band, both of which have the potential to contribute to the observed colorimetric trend[22,23]. Evidence for the former mechanism comes from variable Mn doping experiments, wherein the average Er-Mn separation is decreased synthetically by increasing the Mn doping fraction from 0 to 40%, and causes a monotonic increase (and decrease) in Er red (and green) emission[21]. In other pressure-dependent studies on crystal nanoparticles with constant Mn dopant concentration, the $^4T_1 \rightarrow {^6A_1}$ transition peak redshifted with higher pressures by as much as 33 meV/GPa and its FWHM increased by as much as 10 nm/GPa[22,23]. This pressure-induced greater Er-Mn spectral overlap could also increase the efficiency of energy transfer from the Er green state to Mn and back to the Er red state. However, these mechanisms presume that mechanical perturbations are sufficiently large to strain the rigid ceramic lattice of the nanoparticle core, which may be greater magnitude than perturbations typically observed in biological contexts[13]. Indeed, at biologically relevant pressures of tens of megapascals (or lower), the redshifting and broadening of the Mn $^4T_1 \rightarrow {^6A_1}$ emission is likely to be trivial in enhancing Er red to green emission ratio.

An alternative route to couple pressure and Er-Mn energy transfer takes advantage of energy migration within a heteroepitaxial architecture. In this scheme, Er doped into the nanoparticle core is spatially segregated from Mn in the shell and surrounding polymer matrix. Er-Mn interionic energy transfer has been shown to be highly efficient in UCNPs and fluoride glasses[24,25], with a critical radius of energy transfer on the order of 2 nanometers[26]. Hence, Er emitters on the outer layer of the UCNP core should still have some level of interaction with Mn ions in a 1.5 nm thick UCNP shell. Mn-Mn ion energy transfer in oxide ceramics has been shown to have a critical radius of energy transfer of about 2 nm as well[27]. Feasibly, this would enable Er within the core to couple energetically to Mn within the polymer matrix, which has significantly larger compliance than the nanoparticle lattice. We hypothesize that local densification of Mn ions adjacent to the nanoparticle surface via increased hydrostatic pressure within the polymer matrix would enhance the rate of Mn-Mn energy transfer from the shell to the matrix, thereby siphoning more energy from the green-emitting state of Er dopants within the core. This hypothesized energy transfer route is similar to energy migration mechanisms previously observed in UCNPs, which have demonstrated core-to-shell energy migration from other lanthanide emitter ions (e.g. $Tm^{3+}$) to Mn and also have specifically observed relatively intensified red Er emission in the presence of Mn[20,7,28,29]. Through this potential mechanism, the enhanced elasticity of the polymer matrix could enable smaller stresses to transduce a given colorimetric change, effectively increasing the composite's optical force response. Furthermore, cross-linking provides a relatively straightforward way to tune matrix compliance, and therefore mechanosensitivity.

To optimize composite colorimetric force sensitivity, we investigated two UCNP-polymer composite architectures tailored toward each hypothesized pathway of Er-Mn energy transfer. In the first design, $Er^{3+}$ and $Mn^{2+}$ are both incorporated into the UCNP core crystal lattice and their average interionic distance changes due to external applied force straining the crystal lattice. These UCNPs have a SrLuF:Yb, Er, Mn @ SrYF architecture that encapsulates the Er and Mn together in the core (Figure 1a). In the second design, Er emitters are near the outer core layer of the UCNP and Mn ions are incorporated into the shell, with the option of additional Mn ions crosslinking the surrounding polymer matrix. In this modality, the density of Mn acceptors in the UCNP shell within the immediate proximity of the UCNP core likely increases as the UCNP lattice is compressed. When the polymer matrix is compressed, crosslinking ion density at the UCNP surface is further increased due to polymer compression. These UCNPs have a SrYbF @ SrYbF:$Er_{0.20}$ @ SrYF:$Mn_{0.10}$ architecture that promotes Er placement closer to the nanoparticle surface but still includes a thin (~1.5 nm thickness) passivating shell to maintain UCNP brightness and facilitate energy transfer from the core to external Mn ions (Figure 1b).

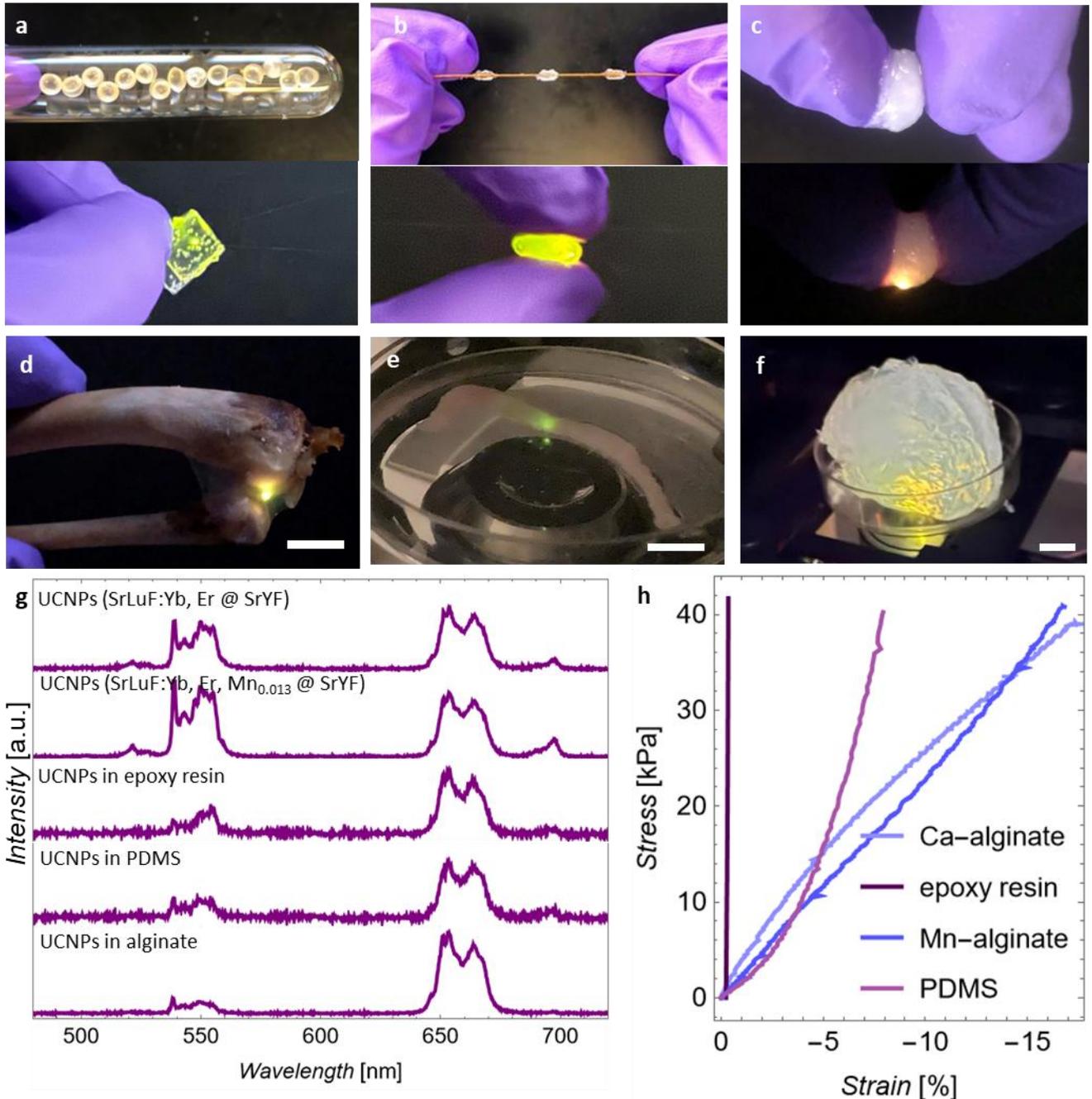

Figure 2. UCNPs incorporated into epoxy, PDMS, and alginates. (a) Epoxy resin beads, (b) PDMS pellets, and (c) alginate hydrogel, without (top) and with (bottom) 980 nm illumination. (d) Epoxy-UCNP composite applied in the joint of a chicken wing bone. (e) PDMS-UCNP composite applied in the lumen of a mouse colon. (f) Mn-alginate-UCNP composite in an agarose phantom brain. Scale bars for (b-d) are 1 cm. (g) Example spectra of UCNPs, bare and in various polymer composites. (h) Compressive stress-strain curves for various polymers.

*Architecture 1: Emitter and d-metal incorporation in the UCNP core*

SrLuF:Yb, Er, Mn @ SrYF cubic phase UCNPs of side length 10.1 ± 2.1 nm (n = 300) were synthesized via thermal decomposition and hot injection as described in the Methods. Average size distributions for core and core@shell nanoparticles were estimated from Transmission Electron Micrographs (TEM), as illustrated in Figure 1c-d. Based on inductively coupled plasma optical emission spectroscopy (ICP-OES), Mn atomic concentrations in the cores were 0%, 1.3%, and 4.2%, with Yb concentration at 34.2 ± 5.7 at. % and Er concentration at 1.5 ± 0.2 at. % (Table S1). The ratio of Er:Yb within the nanoparticle cores remained relatively constant for the three different Mn concentrations, at 0.0436 +/- 0.0060.

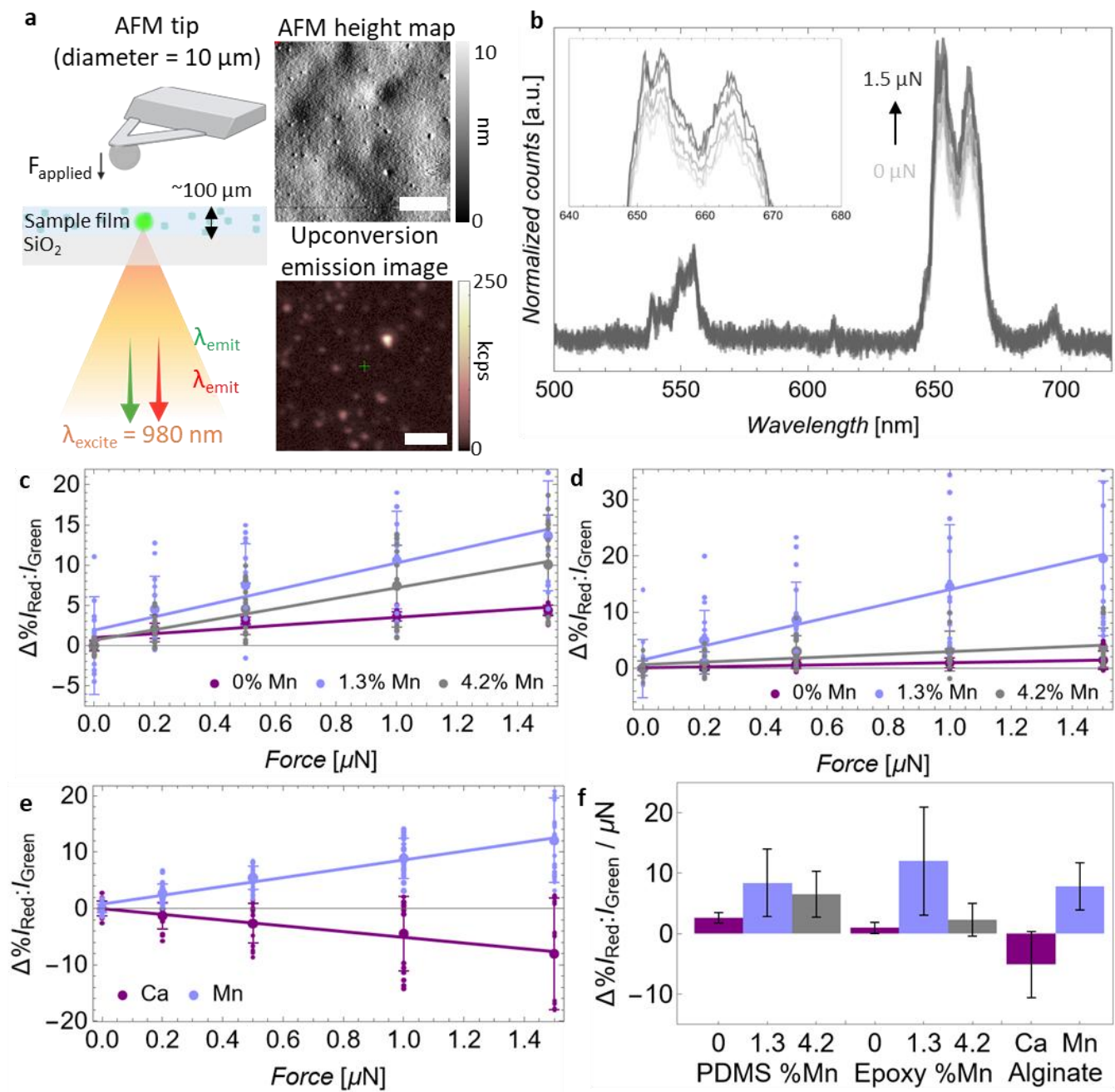

Figure 3. Optical force response calibration of UCNP-polymer films. (a) Schematic of simultaneous atomic force and confocal microscope spectral-force calibration setup. Scale bars 5 μm. (b) Example upconversion emission spectra under different applied forces, normalized to green emission integrated area. (c) Change in red to green emission ratio as a function of applied force for PDMS-UCNP films, (d)

epoxy-UCNP films, (e) alginate-UCNP films. (f) Summary of force sensitivities of different polymer-UCNP films from AFM-confocal calibrations of (c-e). Error bars are standard error of each best fit line.

Synthesized UCNPs were then embedded into polymer hosts of varying stiffness. First, UCNPs were sonicated in dilute acid to remove oleic acid ligands incorporated during synthesis and allow chemical compatibility with polymer precursors, as described in the Methods. Ligand-stripped UCNPs were then mixed in resin precursors for epoxy (Figure 2a) or PDMS (Figure 2b) at a concentration of at 1 mg/mL (see Methods). Baseline optical spectra of the UCNPs with and without Mn-doping, as well as the Mn-doped UCNPs in the different polymer matrices when uncompressed, showed that the presence of the two main red and green emission peaks remained constant, although the red to green emission ratio was greater for the polymer-UCNP composites compared to the bare UCNPs (Figure 2g). The epoxy resin elastic modulus was measured using rheology as ~483 kPa (Figure 2h), making it relevant as a cartilage mimic (see chicken wing joint in Figure 2d)[30]. The PDMS matrix had a measured elastic modulus of ~8 kPa, making it compatible with forces exerted by organ tissues such as the gastrointestinal tract (see mouse colon in Figure 2e)[31].

To characterize the optical force response of the composite films, we used simultaneous confocal-AFM to exert force over a range of 0 to 1.5 µN on the sample while collecting spectral output (Figure 3a). In this technique, we first scanned the sample topography with AFM and upconversion luminescence with 980 nm confocal microscope excitation. We used these two spatial maps to colocalize the AFM tip and confocal laser spot on the sample, then collected a series of spectra at different applied forces. The red and green peak areas of these spectral outputs were used to calculate the integrated red to green emission ratio, a metric that has been previously shown to be more robust to sample or source fluctuations compared to absolute emission intensity[19]. The representative spectra in Figure 3b exemplify the force dependent increase in the integrated red emission relative to green emission. The trend and magnitude of this force dependence is consistent with that previously reported for polystyrene-embedded $NaY_{0.8}Yb_{0.18}Er_{0.02}F_4@NaYF_4$[13]. In the absence of Mn doping, resonant energy transfer exclusively occurs between lanthanide donor-acceptor pairs. For both Er-Er and Er-Yb ion pairs, increased rates of cross relaxation have been shown to preferentially quench the green-emitting Er states compared to the red-emitting states[13]. In alignment, strain-induced reductions in average lanthanide donor-acceptor pair separation induce an increase in $\Delta\%I_{Red}:I_{Green}$. However, as was observed in Casar et al, the ratiometric response sensitivity is on the order of 1% per microNewton[13].

By increasing the Mn dopant concentration, we are able to further sensitize the ratiometric emission response to force. As shown in Figure 3c, non-zero manganese concentrations lead to higher force sensitivity. Interestingly, this effect is not monotonic; a doping concentration of 1.3% Mn yields a sensitivity of 8.4 $\Delta\%I_{Red}:I_{Green}$ per microNewton, while a concentration of 4.2% Mn in the UCNP core yields a sensitivity of 6.5 $\Delta\%I_{Red}:I_{Green}$. This same trend was observed in the stiffer epoxy composite as well. 1.3% Mn doping exhibited a sensitivity of 12 $\Delta\%I_{Red}:I_{Green}$ per microNewton, 4.2% Mn doping had a sensitivity of 2.3 $\Delta\%I_{Red}:I_{Green}$, and no manganese doping had a sensitivity of 0.95 $\Delta\%I_{Red}:I_{Green}$ (Figure 3d). These data indicate that manganese doping increases the ratiometric force sensitivity of UCNPs embedded in polymeric hosts, a finding that corroborates similar results acquired in high pressure diamond anvil cells[19]. Moreover, there appears to be an optimal Mn concentration beyond which the sensitivity is diminished, a trend similarly observed by Lay et al[19]. One explanation for this is that excess Mn:Er allows Mn-Mn cross relaxation to dominate over Er-Mn cross relaxation, diminishing the red-selective emission enhancement effect. Expanding on the confocal-AFM results of Casar et al, these data also indicate that host matrix stiffness is a key factor in tuning composite mechanosensitivity. Across all Mn core concentrations, the stiffer epoxy resin matrix exhibited a tenfold greater $\Delta\%I_{Red}:I_{Green}$ per microNewton slope than the PDMS matrix. This suggests that a less deformable polymer host is necessary to maximize force sensitivity in architectures relying on the compression of the nanoparticle lattice.

*Architecture 2: Emitter and d-metal compartmentalization to core, shell, and polymer matrix*

Next, we investigated the effects of partitioning Er donors and Mn acceptors on ratiometric force sensitivity. We chose alginate hydrogels as the surrounding polymer matrix due to their ability to be ionically crosslinked. We synthesized SrYbF @ SrYbF:$Er_{0.20}$ @ SrYF:$Mn_{0.10}$ cubic phase UCNPs of side length 15.5 +/- 3.1 nm (n = 100) via thermal decomposition and hot injection as described in the Methods. These UCNPs were then incorporated into alginate hydrogels crosslinked with either $Ca^{2+}$ or $Mn^{2+}$ (Figure 2c), and the elastic moduli of both hydrated alginates were measured to be ~2 kPa (Figure 2h). This stiffness is a good match for mechanosensitive imaging in hydrogel phantom brain experiments (Figure 2f).

The UCNP-alginates were also evaluated for spectral force sensitivity using simultaneous AFM-confocal microscopy. We found that the $\Delta\%I_{Red}:I_{Green}$ in the UCNP-Mn-alginate composite increased with applied force, while surprisingly the $\Delta\%I_{Red}:I_{Green}$ slightly decreased for the UCNP-Ca-alginate (Figure 3e). This may be explained by the effects of alginate hydrogel emission peaks spectrally shifting with pressure. The Raman signature of Ca-alginate, which has a broad hydroxyl stretch mode in the 3000-3750 $cm^{-1}$ range[32,33]. In lanthanide fluoride crystal lattices, Er emitters have a green emission-quenching transition ~3150 $cm^{-1}$ and a red emission-quenching transition ~2850 $cm^{-1}$[34]. As water in the hydrogel is compressed, the Raman signature of -OH peaks redshift to lower wavenumbers, lessening overlap with the green-quenching Er transition and increasing overlap with the red-quenching transition and thus decreasing red to green emission ratio[35]. To counteract this effect, we crosslink the alginate hydrogel with $Mn^{2+}$ instead of $Ca^{2+}$ to increase the density of $Mn^{2+}$ acceptors near UCNP surfaces as the hydrogel is compressed, thus increasing the probability of Er-Mn red-enhancing energy transfer. While the slope of $\Delta\%I_{Red}:I_{Green}$ per microNewton for the UCNP-Mn-alginate is comparable to that of the Mn-doped UCNPs embedded in PDMS, UCNP-Mn-alginate still shows slightly less colorimetric change than 1.3% Mn UCNPs embedded in epoxy resin (Figure 3f). Overall, these results suggest that d-metal crosslinking in the alginate polymer matrix is a viable alternative to d-metal doping within the UCNP in terms of colorimetric response to force, and this colorimetric response is highly dependent on the crosslinker ion species used in the alginate hydrogels. This may aid in designing UCNP-polymer force-sensing composites for cases where more flexible and less rigid polymers are more applicable for mechanical matching to soft tissues, for example in probing the brain or skin patches for wound healing.

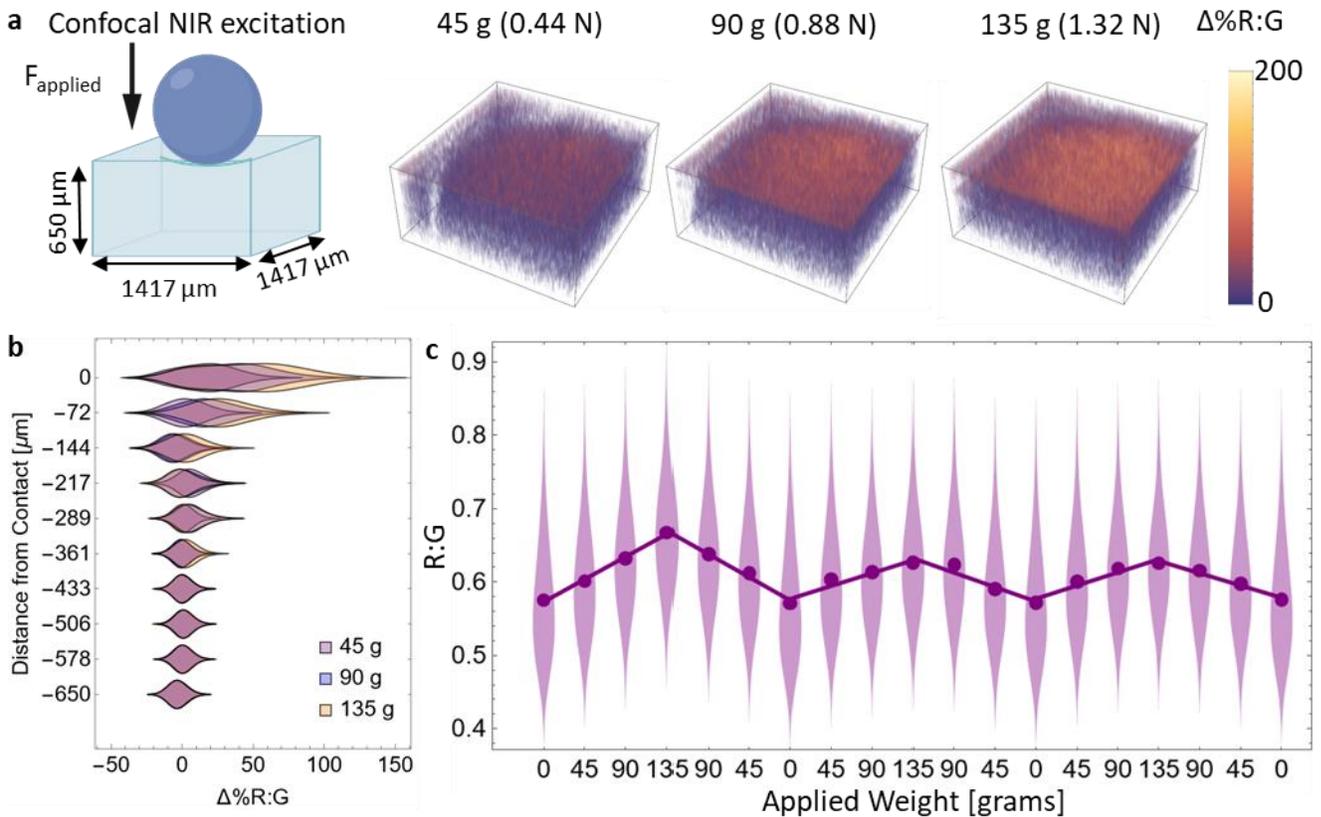

Figure 4. (a) Experimental setup schematic for weighted sphere on epoxy-UCNP composite, with Z-stack images of percent change in red to green emission ratio for bead weights of 45 g, 90 g, and 135 g. (b) Violin plots of percent change in red to green ratio across all pixels in the associated images in (a) as a function of image plane distance from the contact surface between the sphere and composite. (c) Red to green emission ratio distribution at contact surface between the bead and composite for 3 cycles of loading and unloading sphere weight.

*UCNP-polymer composite force response in a simple geometric setup*

        Following from the single-ramp AFM-confocal calibrations on the UCNP-polymer films above, we chose the composite with the greatest optical force sensitivity (epoxy resin embedded with SrLuF:$Yb_{0.28}Er_{0.025}Mn_{0.013}$ @ SrYF) and demonstrated its force mapping capability in a simple sphere-on-plane contact geometry, as shown in Figure 4a. Using a custom-printed mount on a confocal microscope stage, a steel ball bearing was stabilized on top of a 1 mm thick composite slab and loaded with tungsten weights to apply compressive force to the composite. At each weighted state, a 650 μm deep Z-stack of 10 images beginning at the contact surface plane was taken and compared to the baseline unweighted state to create spatial maps of the percent change in red to green emission ratio (Δ%R:G), shown in Figure 4a. Qualitatively, and as expected, the top image plane has higher Δ%R:G with greater applied weight on the steel sphere. The Δ%R:G distributions in each image plane of these Z-stacks were plotted as a function of distance from the top plane of contact surface, for different applied weights (Figure 4b). More than 350 μm below the contact surface, the Δ%R:G distributions appear quite similar regardless of applied weight. Closer to the contact surface, Δ%R:G distributions skew upward for greater applied weight, and at the contact surface, the median Δ%R:G increases from 16 for 45 grams weight, to 34 for 90 grams weight, to 47 for 135 grams weight. This suggests that the stress differences from the weighted steel sphere on the epoxy-UCNP slab are most drastic at the plane of contact and dissipate

further out, in accordance with the Hertz model of contact mechanics[36]. To demonstrate the cyclability of the epoxy-UCNP composite optical force response, the absolute red to green emission ratio (R:G) distribution was measured at the plane of contact, with 3 cycles of loading and deloading weight on the steel sphere (Figure 4c). The median R:G scales from 0.56 to 0.66 on the first cycle load ramp, 0.56 to 0.62 on the second, and 0.57 to 0.62 on the third, consistently increasing and decreasing with load and deload, respectively. Although there is a slight decrease in the range of the median R:G with subsequent cycles, this can be expected from a viscoelastic polymer matrix. These data confirm the polymer-UCNP composite's utility as an optical force-mapping probe.

*UCNP-polymer composite force sensing on wide field imaging platform*

To spatially map forces in the epoxy-UCNP composite, we use a wide field imaging setup with 980 nm illumination and 3 mW input laser power. This imaging system, rather than giving a spectral readout, has imaging channels for red intensity and green intensity, and the red to green ratio is calculated by overlaying these images and calculating their relative intensities (R:G) (Figure 5a). Using a 4 mm tall and 5 mm diameter cylinder of epoxy-UCNP composite, we adjusted the uniaxial compressive stress on the cylindrical sample by stacking calibration weights on top of it while imaging the red and green intensities of the upconversion emission. To process the red to green intensity of the images, we plot the red to green ratio as a function of uniaxially applied compressive stress on epoxy-UCNP composite (Figure S9). This calibration curve gives a Δ%R:G of 0.515% per kPa of uniaxial compressive stress, or about 10.1 Δ%R:G per Newton

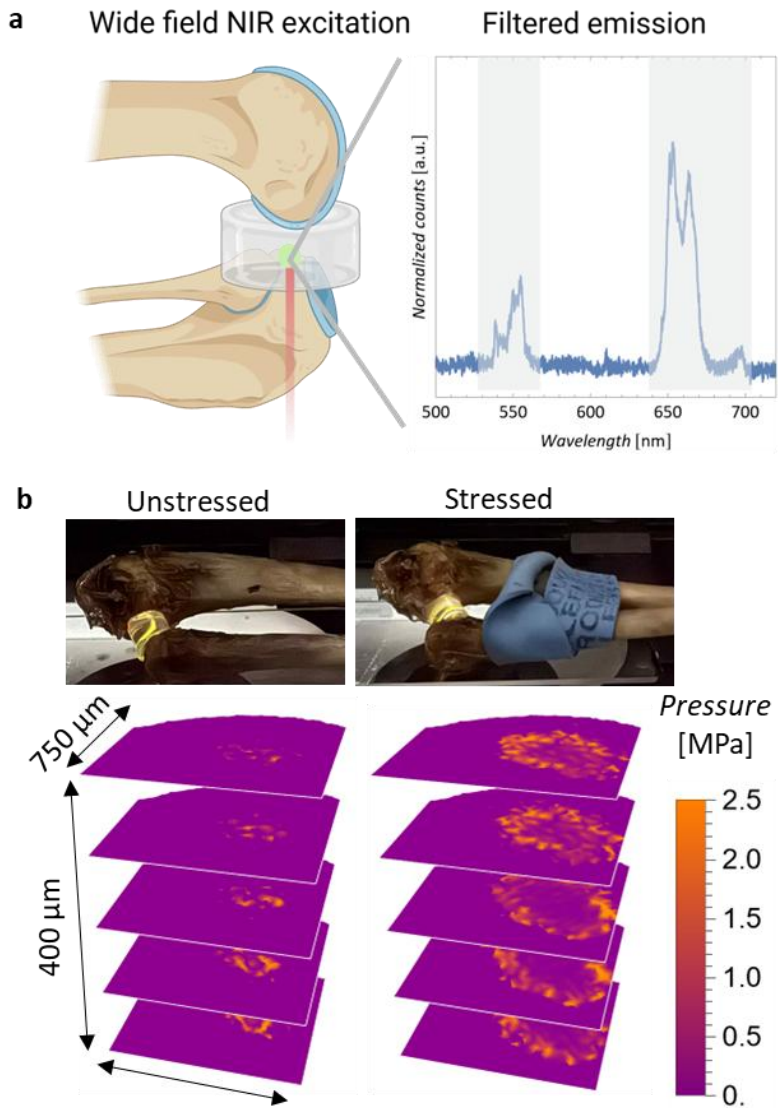

Figure 5. (a) Schematic of imaging workflow, from sample illumination on wide field microscope to filtered emission images. (b) 3D experimentally measured stress from red to green emission ratio maps of bulk epoxy-UCNP composite between the joint of a chicken wing bone, in unstressed and stressed states (without and with rubber band).

To demonstrate the epoxy-UCNP composite's utility as a force sensor in an articular cartilage model, we place a piece of composite between the bones in the joint of a chicken wing and measure the red to green ratio with and without the stress of a rubber band around the wing bones (Figure 5b). We find that from the unstressed state to the stressed state, peak measured local $\Delta\%I_{Red}:I_{Green}$ corresponded to peak pressure of about 2.5 MPa according to the calibration curve in Figure S9. Measurements were taken at the center of the composite cylinder, where forces would be expected to be more uniform, rather than at the surface point of contact with the bone, where force changes would be expected to be most drastic. Similar measurements using a epoxy-UCNP composite coating on the bone joint instead of a thick composite cylinder can be found in Figure S10. Peak forces in the human knee have been estimated to reach about 3 times body weight[37], and given a global average human body weight of 60 kg[38], an estimated peak force of 1766 N in the human knee spread over several square centimeters of tissue area gives up to tens of MPa of peak pressure in the knee. If incorporated into knee articular cartilage prosthetics, this epoxy-UCNP composite could feasibly serve as a minimally invasive pressure-sensitive probe that resolves MPa pressures. However, extensive studies on and improvements to the material's biodegradation hazards and *in vivo* durability would be needed before such technology could be safely put into live animal models or clinical practice.

## 3. Conclusions

We explored the optical force sensitivity of UCNPs with varying core-shell architectures and Mn doping concentrations embedded in PDMS, epoxy resin, and alginate hydrogel. Among all the combinations of UCNP and polymer architectures we tested, SrLuF:Yb$_{0.28}$Er$_{0.025}$Mn$_{0.013}$ @ SrYF UCNPs embedded in epoxy resin showed the greatest sensitivity (Δ%I$_{Red}$:I$_{Green}$ per microNewton) as determined by simultaneous AFM and confocal microscopy measurements. We used this epoxy-UCNP composite to spatially map forces in a sphere-on-plane contact geometry on a confocal microscope and demonstrated that it had a consistently cyclable optical force response. We also characterized the Δ%I$_{Red}$:I$_{Green}$ of the bulk epoxy-UCNP composite on a wide field microscope and demonstrated its utility as a pressure sensor in the joint of a chicken wing. With several optimization tuning knobs in place, ranging from precise modification of UCNP composition and architecture to variable polymer functional and mechanical properties and probe geometries, these composite mechanosensors can be adapted to optimize sensitivity at different force scales. Additionally, the delivery method can be tailored for a targeted biological system or model.

Looking forward, we envision a new generation of polymer-UCNP force sensors that could enable real-time, minimally-invasive force detection in biological systems. These sensors are especially well suited to force detection at tissue interfaces or within lumens. Future investigation into alternative d-metal dopants such as chromium, cobalt, or iron into the UCNP cores may lead to enhanced spectral overlap with erbium energy levels and improved emission color modulation. Additionally, optimizing the polymer matrix through controlled crosslinking density and d-metal concentration may help to refine energy migration and transfer dynamics between the UCNP emitters and their external environment. In-depth biocompatibility and durability assessments on future polymer-UCNP composites will be required to support their application in mechanobiology research and implantable diagnostics across animal models and tissues of diverse mechanical properties.

## 4. Methods

*Materials*

Trifluoroacetic acid (TFA, 99%), erbium oxide, ytterbium oxide, manganese(II) acetate tetrahydrate were purchased from Alfa Aesar. Oleic acid (OA, 99%), 1-octadecene (ODE, 90%), oleylamine (OLE, 70%), diethyl ether, strontium carbonate, calcium chloride, yttrium acetate, lutetium oxide, hydrochloric acid, nitric acid, hexanes, cyclohexanes, ethanol, and isopropyl alcohol solvents were purchased from Sigma Aldrich. Sylgard 184 polydimethylsiloxane (PDMS) was purchased from Fisher Scientific. Epoxy resin was a commercially available NicPro brand. Manganese, strontium, lutetium, ytterbium, and erbium standards for inductively coupled plasma optical emission spectroscopy were purchased from Agilent Technologies. Sodium alginate was purchased from Ward's Chemicals. All chemicals were used as purchased without further purification.

*SrLuF:Yb, Er, Mn @ SrYF Upconverting Nanoparticle Synthesis*

Core-shell SrLuF host lattice nanoparticles were synthesized by a thermal decomposition procedure modified from that which has been previously described[12]. Briefly, trifluoroacetic acid precursors were first made from metal oxide or metal acetate salt. 20 mmol of stoichiometric metal (e.g.

20 mmol SrCO$_3$, 10 mmol Lu$_2$O$_3$, 10 mmol Mn-acetate) oxide or acetate salt powder was added to a 250 mL round-bottom three-neck flask. 100 mL distilled water and then 20 mL TFA was added to the flask, which was then heated to 90 °C in an oil bath and stirred at 500 rpm for at least 30 minutes, or until the solution became transparent. The oil bath temperature was then lowered to 65 °C and the remaining liquid was evaporated overnight until only a white powder of metal-TFA salt was left. Then, 20 mL of OA and 30 mL of ODE were added to the flask. The solution was heated to 120 °C under vacuum, then cycled between N$_2$ gas and vacuum three times to remove residual water and oxygen before cooling back down to room temperature. This procedure gave a TFA solution with 1 mL OA and 1.5 mL ODE per mmol of metal.

To make SrLuF:Yb, Er, Mn cores, 0.5 mmol of Sr-TFA precursor and a total of 0.5 mmol lanthanide TFA precursor were mixed the day before synthesis. The same day of synthesis, this TFA precursor mixture was stirred in a 50 mL 3-neck round bottom flask at 300 rpm. Then, accounting for the OA and ODE already present in the precursor mixture, OA was added to a total of 3 mL and ODE added to a total of 9.5 mL, along with 1 mL oleylamine. Stirring rate was increased to 1000 rpm, vacuum was pulled on the flask, and the mixture was heated to 120 C using a heating mantle. The flask was then cycled between vacuum and N$_2$ inert gas three times, then heated quickly to 305 °C under inert gas with added glass wool insulation to the upper part of the flask. The mixture was held at 305 °C for 1 hour, then quickly cooled to room temperature using a heat gun on the cool air setting. Particles were precipitated by adding about 7.5 mL ethanol to the solution and washed by centrifugation at 3000 g for 5 minutes to obtain a solid whitish pellet. Supernatant was discarded and the pellet was redispersed in about 5 mL hexanes, followed by adding about 25 mL ethanol, and centrifuged again at 3000 g for 5 minutes. The resulting pellet was redispersed in 5 mL cyclohexanes to obtain a UCNP concentration of about 0.1 mmol per mL.

For the inert shell, Y-oleate precursor was prepared by adding Y-acetate, OA, and ODE to a 3-neck round bottom flask in a ratio of 1 mmol:2mL:2mL, respectively. The mixture was put under vacuum and heated to 120 °C, cycled between vacuum and N$_2$ inert gas 3 times until solids were all dissolved, then cooled to room temperature.

Cores were shelled with SrYF by hot injection. The day before shelling, 1 mmol Sr-TFA and 1 mmol Y-oleate precursors were mixed together. To a 50 mL 3-neck round bottom flask, 0.1 mmol of SrLuF:Yb, Er, Mn cores in cyclohexanes (1 mL of 1 mmol/10 mL concentration), 3 mL OA, and 6 mL ODE were added and stirred at 500 rpm at 65 °C with two open flask necks to evaporate off cyclohexanes. Meanwhile, the pre-mixed Sr-TFA and Y-oleate solution was prepared in a 12 mL syringe with a 3-inch long hypodermic needle and a syringe pump was set to dispense at 2.2 mL/hr. The stirring in the 3-neck flask was increased to 1000 rpm, put under vacuum, and heated to 120 °C, then cycled between vacuum and N$_2$ inert gas 3 times. The flask was then set to heat to 300 °C and the syringe was inserted through a rubber septum into the flask so that the needle was positioned above the solution but not touching. Glass wool insulation was added to the upper part of the flask. When the flask reached 270 °C, syringe pump injection was started. Once all shelling solution was injected, the syringe pump was stopped, the flask was taken off heat and allowed to slowly cool to room temperature. Particles were then washed using the same procedure as washing the cores. Final shelled particles were redispersed in 4 mL hexanes.

*SrYbF @ SrYbF:Er$_{0.20}$ @ SrYF:Mn$_{0.10}$ Upconverting Nanoparticle Synthesis*

To synthesize nanoparticles that better facilitate energy transfer between the Er ions and external Mn ions, first SrYbF core nanoparticles were made by thermal decomposition. As described previously, Sr-TFA and Yb-TFA were first made from SrCO$_3$ and Yb$_2$O$_3$ salts.

SrYbF cores were then grown with an active SrYbF:Er$_{0.20}$ shell by hot injection. A 1:0.8:0.2 Sr:Yb:Er TFA precursor mixture was made the day before shelling. In a 50-mL three-neck round-bottom flask, 1 mL of the SrYbF cores dispersed in hexanes, 3 mL oleic acid, and 6 mL octadecene were added, heated to 65 C to evaporate the hexanes, and stirred at 500 rpm. A syringe pump was prepped with the SrYbEr-TFA shell precursor solution and set to dispense at a rate of 2.2 mL/hr. The remaining shelling and particle washing procedure was the same as the SrYF shelling procedure as described in the previous section. Particles were redispersed in 4 mL hexanes.

SrYbF @ SrYbF:$Er_{0.20}$ nanoparticles were additionally shelled with a thin layer of SrYF:$Mn_{0.10}$, an inert shell with some Mn ions to facilitate energy transfer to external Mn ions. The day before shelling, 0.5 mmol Sr-TFA precursor was mixed with 0.45 mmol Y-oleate and 0.05 mmol Mn-TFA precursors. The day of shelling, 0.05 mmol of SrYbF @ SrYbF:$Er_{0.20}$ cores in hexanes (2 mL of 0.1 mmol/4 mL concentration) was added to a 50 mL 3-neck round bottom flask along with 3 mL OA and 6 mL ODE, then stirred at 500 rpm at 65 °C with two open flask necks. The SrYF:$Mn_{0.10}$ shelling solution was prepared in a 5 mL syringe on a syringe pump set to dispense at 2.12 mL/hr. The solution in the flask was put under vacuum, increased stirring rate to 1000 rpm, and heated to 120 °C. The remaining shelling and particle washing procedure was as described in the previous section. Particles were redispersed in 4 mL hexanes.

*Nanoparticle Ligand-stripping*

Nanoparticles were stripped of oleic acid ligands via a previously described procedure[39]. Briefly, nanoparticles dispersed in hexanes were air-dried and massed in a glass vial. For every 10 mg of dried nanoparticles, 1 mL of 0.04 M HCl in 80% EtOH and 20% $H_2O$ solution was added to the vial. The vial contents were sonicated for at least 30 minutes and then poured into a 60 mL separatory funnel. The vial was rinsed with about 10 mL $H_2O$, which was also added to the funnel. About 30 mL diethyl ether was added to the funnel, which was then stoppered and shaken vigorously to mix the contents. The stopper was removed and the funnel contents were allowed to separate, with oleic acid remaining in the upper diethyl ether layer and nanoparticles in the denser aqueous phase. The aqueous phase was collected and the diethyl ether discarded. The funnel was rinsed twice with EtOH and the aqueous phase was returned to it. About 15 mL diethyl ether was added to the funnel again, and the shaking and aqueous phase collection process was repeated. The aqueous phase was transferred to a 50 mL Falcon tube and centrifuged at 8000 g for 10 minutes. The final pellet of ligand-stripped nanoparticles was redispersed in either hexanes, IPA, or water and sonicated thoroughly.

*Nanoparticle Morphological Characterization*

Transmission electron microscope images of nanoparticles were taken on a FEI Tecnai G2 F20 X-TWIN with a field-emission gun at 200 kV operating voltage and a Gatan SC200 camera.

*Nanoparticle Inductively-Coupled Plasma Optical Emission Spectrometry (ICP-OES)*

ICP-OES was conducted on the UCNP cores with a Thermo Scientific ICAP 6300 Duo View Spectrometer with a solid state CID detector. Samples were prepared by dissolving about 2 mg of dried UCNP cores in 10 mL of 5% nitric acid. Purchased standards were diluted with 5% nitric acid to concentrations expected to match those of the UCNP samples by 0x, 0.5x, 1x, 2x, and 4x, with an extra 2x concentration standard for quality control. UCNP core element concentrations were interpolated from the standards' concentration curves, and UCNP core manganese atomic concentration was calculated as a ratio of manganese ICP concentration to strontium ICP concentration.

*Polymer Morphological Characterization*

Scanning electron microscope images of polymer matrices were taken on a Hitachi desktop SEM with 5 kV operating voltage and mixed backscatter and secondary electron signal.

*UCNP Polymer Incorporation*

For PDMS incorporation, ligand-stripped UCNPs dispersed in hexanes were added in equal volume to a Sylgard 184 PDMS mixture (10:1 resin:curing agent by volume) and mixed thoroughly. For epoxy resin incorporation, ligand-stripped UCNPs dispersed in IPA were added in equal volume to an epoxy resin mixture (1:1 resin:hardener by volume) and mixed thoroughly. To form thin films for opto-mechanical characterization, 200 μL of the polymer-UCNP mixtures were spin-coated onto glass coverslips directly after mixing at 2000 rpm for 3 minutes. PDMS-UCNP samples were then cured in an oven furnace at 40 °C for 2 hours, and epoxy resin-UCNP samples were cured on a hotplate at 60 °C overnight.

For alginate incorporation, first a 3 % (wt/vol) Na-alginate solution was made by adding 600 mg sodium alginate powder to 20 mL distilled water and stirring at 70 °C for 2-3 hours until powder is completely dissolved. Ligand-stripped UCNPs dispersed in water were added in equal volume to the 3 % (wt/vol) aqueous Na-alginate solution and vortexed to mix thoroughly. To form thin films for opto-mechanical characterization, 100 μL of the alginate-UCNP mixture was spin-coated onto glass coverslips directly after mixing at 1000 rpm for 2 minutes. The coverslip surface was then covered with $MnCl_2$ (1 M) aqueous solution or 1 M $CaCl_2$ aqueous solution and alginate was allowed to cure overnight, forming Mn-alginate or Ca-alginate respectively.

*Opto-mechanical UCNP-polymer Film Characterization*

A dual atomic force and confocal microscope setup was used to characterize polymer-UCNP opto-mechanical response. The setup (Figure S4) was built using a Zeiss Axio Observer.Z1m confocal microscope with an input 980 nm excitation laser, with an Asylum Research MF-2PD atomic force microscope mounted on the stage. The emission signal output was split and routed to an Excelitas Technologies Single Photon Counting Module avalanche photodiode and a Princeton Instruments IsoPlane SCT320 spectrometer.

A typical trial of simultaneous atomic force and confocal microscopy opto-mechanical characterization on a UCNP-embedded polymer film is as follows. First, a 10 μm diameter SiO sphere AFM tip was calibrated using Asylum Research's Igor software over a clean glass coverslip. The clean coverslip was then switched out for a thin polymer-UCNP spin-coated sample and the confocal microscope was used to find the focal height of the sample. For epoxy-UCNP and PDMS-UCNP films, an AC scan of the film topography was taken, and clusters were matched to bright spots on the optical confocal map of upconversion signal to co-locate the AFM tip and confocal microscope excitation laser. For alginate-UCNP films, the sample and tip were immersed in a water droplet and allowed to equilibrate for at least 30 minutes, then the tip was recalibrated in water according to the Asylum Research MF3PD manual. Using the camera mounted in the AFM head, the tip was engaged to 0 nN and the confocal laser spot was laterally co-located with the spherical AFM tip on the sample, then vertically focused to maximize the emission counts. The tip calibration constants were used to calculate deflection voltage values corresponding to specific forces and input into the MacroBuilder software. Three spectra were collected at each force step, increasing the tip force on the sample from 0 to 1500 nN, then the tip was withdrawn completely. This was repeated for each UCNP-polymer combination for at least 5 trials, where trials were included in data analysis if the overall emission counts increased during the force increase, indicating that the confocal laser was truly underneath the AFM tip.

Collected spectra were analyzed in Mathematica (Wolfram, Inc.). Green emission area was integrated from 510 to 570 nm wavelengths and red emission area was integrated from 630 to 710 nm wavelengths. The average R:G at 0 nN for each trial was then set at 0 $\Delta\%I_{Red}:I_{Green}$ and the $\Delta\%I_{Red}:I_{Green}$ for each subsequent step was calculated using the 0 nN R:G as the 0 $\Delta\%I_{Red}:I_{Green}$ reference. $\Delta\%I_{Red}:I_{Green}$ was averaged over all trials for each UCNP-polymer combination.

*Rheological properties of polymers*

Polymer compressive moduli were collected using a TA Instrument ARES-G2 rheometer with a flat parallel 25 mm stainless steel plate. The epoxy resin sample was prepared by mixing a 1:1 ratio of resin to hardener, then curing in a PDMS mold at 60 °C on a hot plate overnight. PDMS samples were prepared by mixing a 10:1 resin to curing agent ratio of Sylgard 184, and then adding an equal volume of hexanes solvent, mixing thoroughly, and curing at 40 °C for 2 hours in an oven furnace. Ca- and Mn-alginate hydrogel samples were prepared by dropping 1 M $CaCl_2$ or 1 M $MnCl_2$ solution to cover an aqueous 3 % (wt/vol) Na-alginate solution, then left overnight to crosslink and henceforth kept in water. Samples were cut to roughly 25mm diameter disks to fit the rheometer sample plate area and axial compression was applied at 20 °C at a Hencky strain rate of -50%/min for 60 s or until axial force reached about 20 N to collect compression stress versus strain curves. Compressive moduli were calculated by linearly fitting the elastic portion of the curves, between axial force from 0 to 20 Newtons.

*Confocal weighted bead geometry force mapping*

Epoxy resin sample was prepared by mixing 1:1:2 ratio of resin:hardener:UCNPs ligand-stripped into IPA solvent and cured in a PDMS mold at 60 °C on a hot plate overnight. A 5 mm diameter steel ball bearing was stabilized on top of the epoxy-UCNP sample using a custom 3D-printed mount on the stage of an inverted Zeiss LSM 780 scanning confocal microscope coupled with a Spectra Physics MaiTai tunable pulsed laser set to 980 nm excitation, with peak pulse power ~675 mW and pump power 12396 mW. Using the Zeiss Zen Black software, laser power was set to 50%, pinhole size set to the maximum 600.6 um, and two detection channels were set to integrate from 509 to 571 nm (UCNP green emission) and 630 to 710 nm (UCNP red emission), with detector gain set to 800 for both respectively. Using a 10x EC Plan Neo air objective (NA = 0.3, WD = 5.2 mm) to focus the Z height at the top plane of the epoxy-UCNP composite in contact with the steel ball, we set that to be the top image focus and collected a Z-stack of 10 images spaced evenly over 650 um total sample thickness. Image XY area was 1417 x 1417 um and 1024 x 1024 pixels. To add downward force to the steel ball, tungsten disk-shaped weights of 45 g each were placed on the mount above the ball up to 135 grams. For each weight, a Z-stack of images was taken with and without (baseline) the calibration weight on the same sample area.

For force cycling demonstration, the tungsten weights were subsequently loaded on and off of the sample and ball mount for 3 complete load-deload cycles, and at each load step, one image scan was taken at the plane of contact between the epoxy-UCNP sample and the steel ball to obtain a red to green emission ratio distribution.

Image red to green ratio analysis was coded in Mathematica. Images were first de-noised with a Gaussian filter of 5 pixels, then the red to green ratio was calculated from the two detection channel intensities and mapped over the entire measurement volume.

*Bulk material wide field image force calibration*

The epoxy resin-UCNP composite was mixed in a 0.5 mL Eppendorf tube and left to cure at 60 °C overnight to create a 5 mm diameter, 4 mm tall epoxy resin-UCNP composite cylinder. This cylinder was placed on a microscope slide on the stage of a Zeiss Axio Observer.Z1m wide field microscope with a Hamamatsu W-View Gemini Image Splitter and Orca Flash 4.0 camera split into two imaging channels, one with a 568 nm long pass filter (red) and one with a 612 nm short pass filter (green). Calibration weights of 0 g, 50 g, 100 g, 200 g, and 500 g were subsequently placed on top of the epoxy-UCNP cylinder, the Z focal height adjusted to the center of the cylinder, and for each weight, 3 images were taken at 5 different XY locations with 100 ms exposure time. The images were analyzed using Mathematica code, using reference background images to remove background noise, a smoothing Gaussian filter of 5, and an intensity threshold of 10 times the background noise standard deviation to remove unilluminated areas outside the beam spot. The calibration weight values and the epoxy-UCNP cylinder cross-sectional area were used to estimate the uniaxial compressive stress on the bulk material, and the red to green ratio was calculated from the red and image intensities.

For mapping the epoxy-UCNP composite cylinder between the chicken wing bones, a similar method was used and the images were processed with the same Mathematica code. To compress the bone joint, a thick rubber band was wrapped zero, two, or three times around the adjacent bones. Images were taken at a focal height range of 0.4 mm in intervals of 0.1 mm with 1 s exposure time around the center of the composite cylinder, and the images were analyzed using the same Mathematica code as for the calibration.

**Acknowledgements**

The authors would like to acknowledge Stanford Nano Shared Facilities, Soft and Hybrid Material Facility, and Cell Sciences Imaging Facility for access and training on characterization equipment. J.A.D. is a Chan Zuckerberg Biohub – San Francisco Investigator. We also acknowledge funding from Q-NEXT, a U.S. Department of Energy Office of Science National Quantum Information Science Research Centers under Award No. DE-FOA-0002253. Additional support is from. the National Science Foundation Graduate Research Fellowships Program (NSF), and the Department of Defense National Defense Science and Engineering Graduate Fellowship (NDSEG).

**Supporting information: Mechanosensitive polymer matrices of biologically-relevant compliance based on upconverting nanoparticles**

Figure S1. Size distributions of UCNPs. (a) Histograms of core and shelled UCNP diameter distributions (n = 100 for all datasets). (b) Averages and standard deviations of core and shelled UCNP diameter distributions (n = 100 for all datasets).

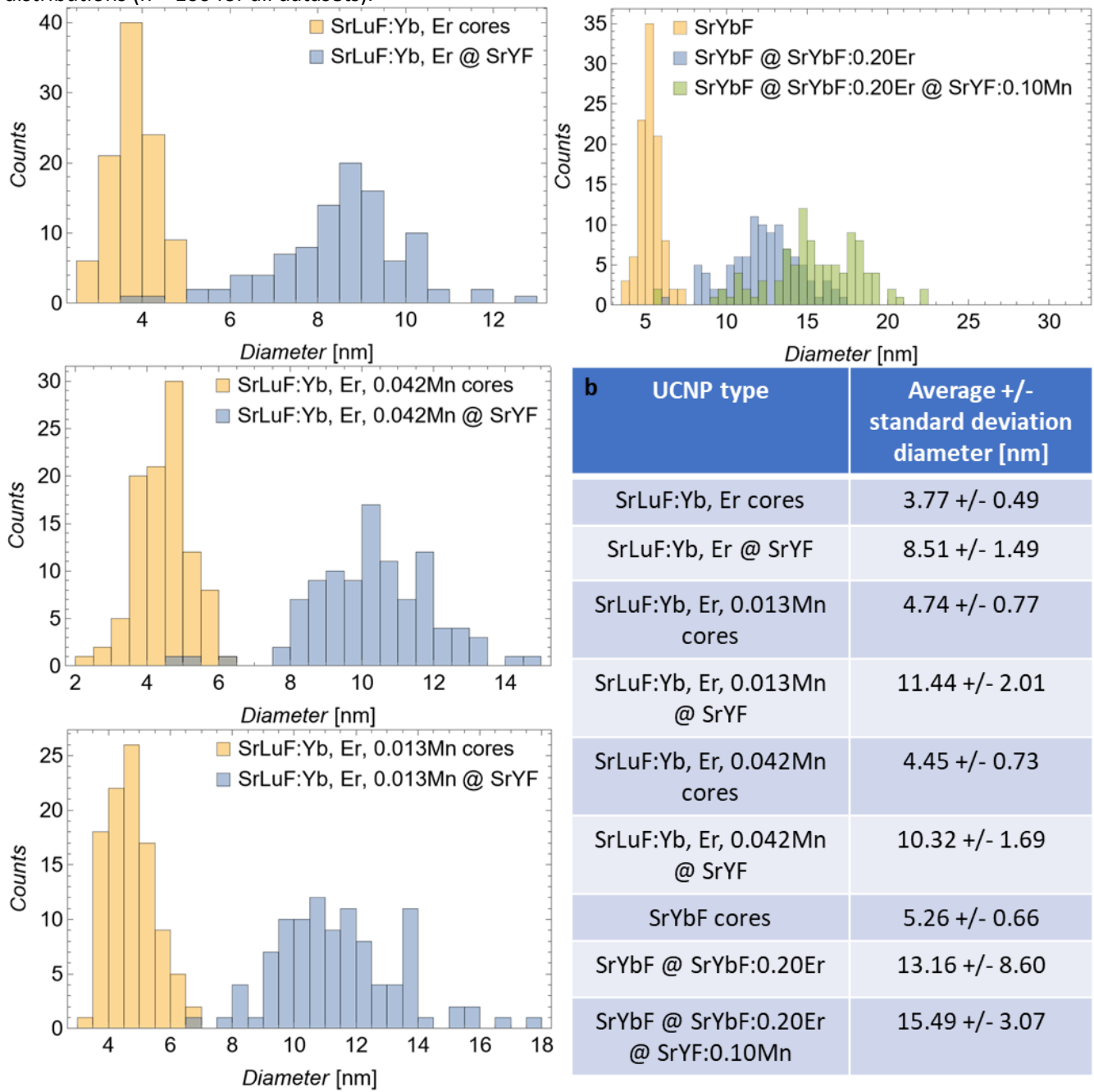

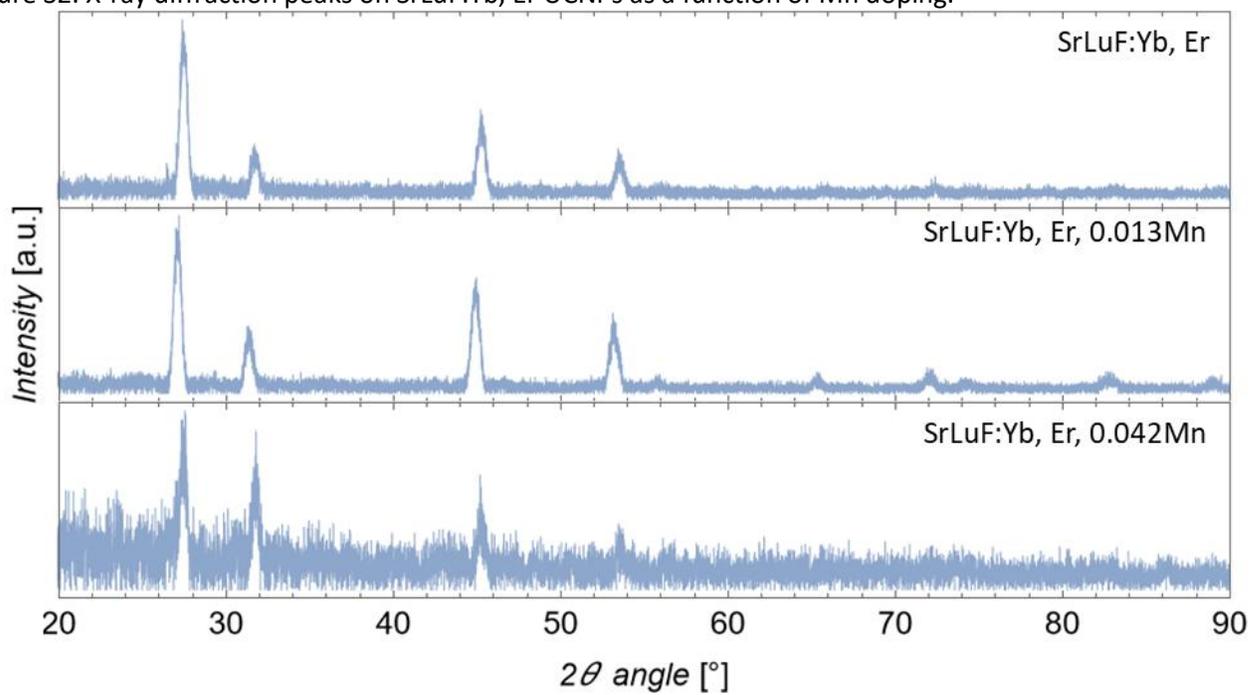

Figure S2. X-ray diffraction peaks on SrLuF:Yb, Er UCNPs as a function of Mn doping.

Figure S3. UCNPs dispersed in hexanes as synthesized, for (a) SrLuF:Yb$_{0.28}$Er$_{0.025}$Mn$_x$ cores, (b) SrLuF:Yb$_{0.28}$Er$_{0.025}$Mn$_x$ @ SrYF, and (c) SrYbF @ SrYbF:Er$_{0.20}$ @ SrYF:Mn$_{0.10}$.

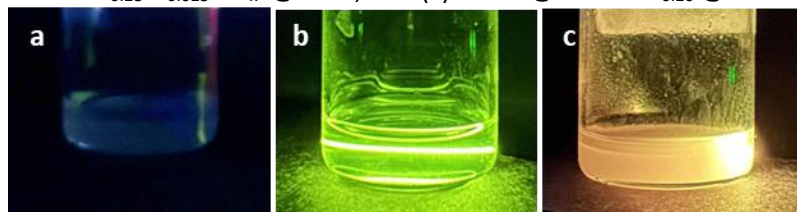

Table S1. Relative concentration values of lanthanide dopants in UCNP cores calculated via inductively coupled plasma optical emission spectroscopy.

| Sample | Er/total Ln | Lu/total Ln | Mn/total Ln | Yb/total Ln |
|---|---|---|---|---|
| SrLuF:Yb, Er | 0.0134 | 0.6227 | 6.6739E-05 | 0.3638 |
| SrLuF:Yb, Er, 0.013Mn | 0.0177 | 0.5852 | 0.0132 | 0.3839 |
| SrLuF:Yb, Er, 0.042Mn | 0.0133 | 0.6676 | 0.0418 | 0.2773 |

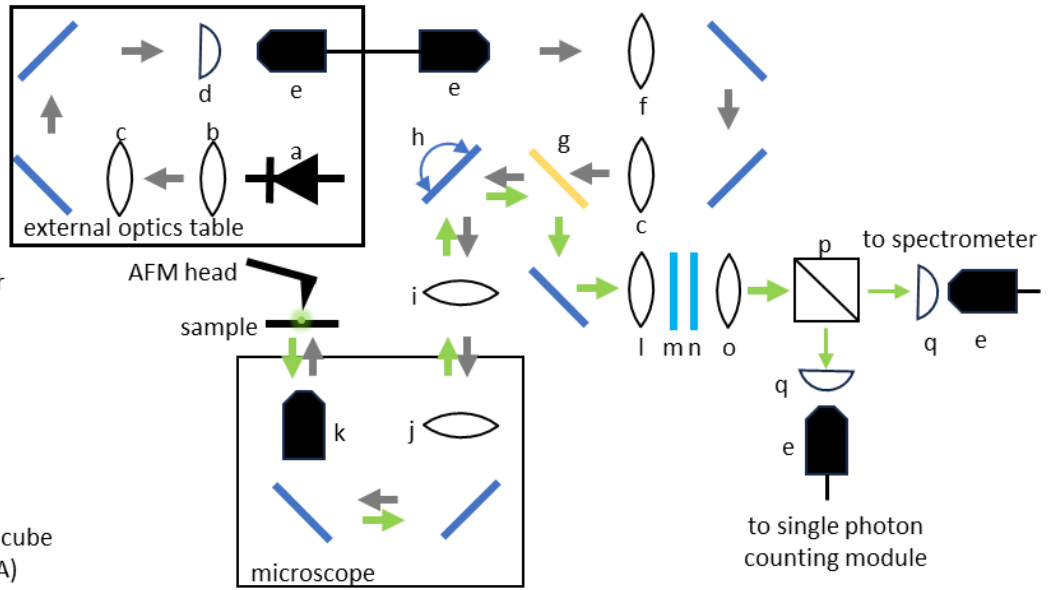

Figure S4. Optical path of dual AFM-confocal microscope. Adapted from Casar *et al.*, *Nature* (2025).

**Key**
a: 980 nm laser
b: 50 mm (B coated)
c: 150 mm (B)
d: 7 mm collimator
e: single mode fiber
f: 60 mm (B)
g: 925 nm dichroic
h: fast scanning mirror
i: 75 mm (AB)
j: 300 mm (uncoated)
k: objective
l: 100 mm (A)
m: 750 shortpass
n: 950 shortpass
o: 60 mm (A)
p: 50:50 beamsplitter cube
q: 11 mm collimator (A)

Figure S5. Fourier Transform Infrared spectroscopy of alginate hydrogels, epoxy resin, and PDMS without nanoparticles. Red and green vertical lines indicate energy shifts that help populate red-emitting and green-emitting energy states in Er$^{3+}$ ions, respectively [(Carnall et al. 1978)].

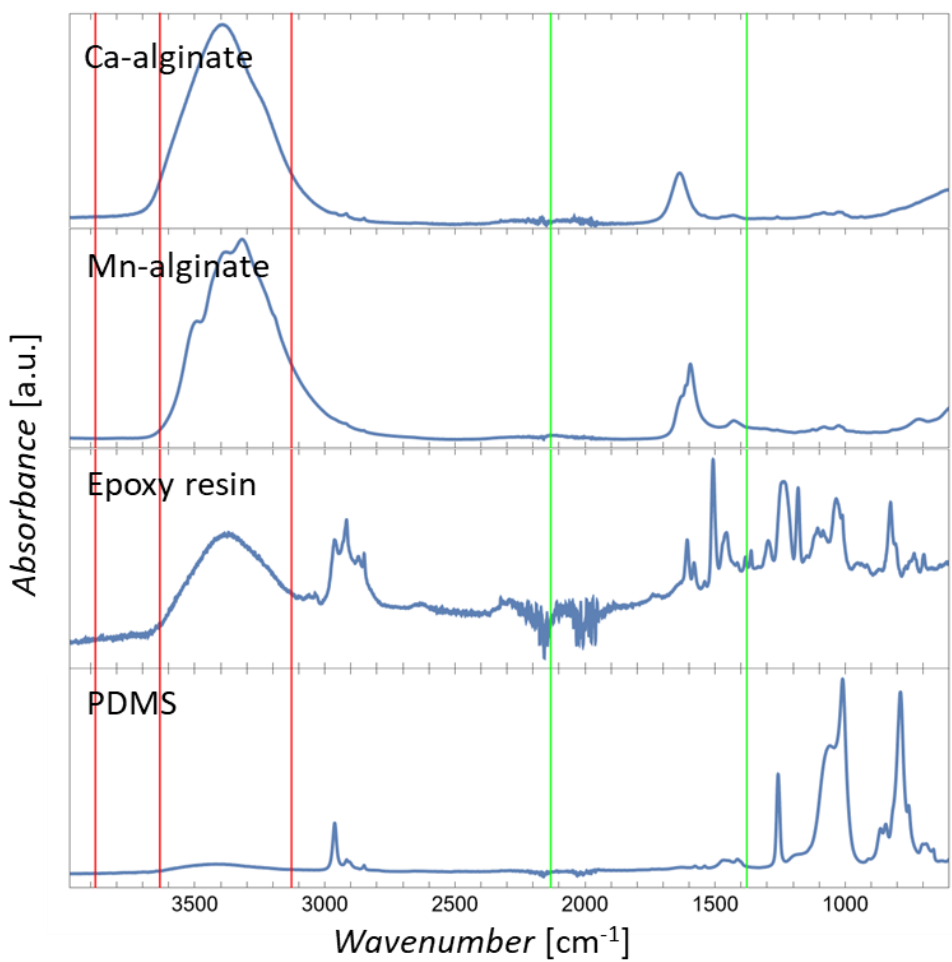

Figure S6. UV-visible spectra of Mn-alginate and Ca-alginate hydrogels. Ca-alginate shows a peak around 1740 nm or 5747 cm$^{-1}$ which could initially couple with the Er$^{3+}$ $^4F_{9/2} \rightarrow {}^4F_{7/2}$ transition around 5260 cm$^{-1}$ [Carnall et al. 1978] (see Figure S7) and help depopulate Er$^{3+}$ green-emitting states and populate red-emitting states, whereas Mn-alginate has a transmission peak at the same wavelength and wavenumber. Upon compression, this Ca-alginate absorption peak might decouple from the Er$^{3+}$ $^4F_{9/2} \rightarrow {}^4F_{7/2}$ energy transition, leading to a downward Δ% $I_{Red}$:$I_{Green}$ trend with increasing compression rather than an upward trend as seen with Mn-alginate (Figure 3e).

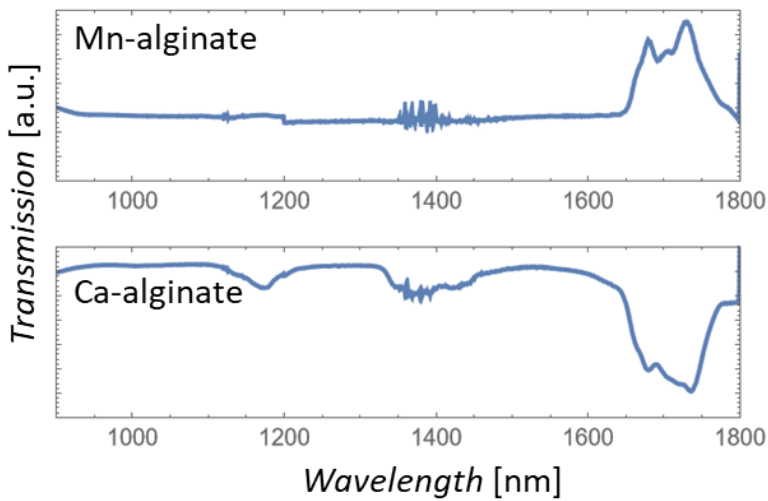

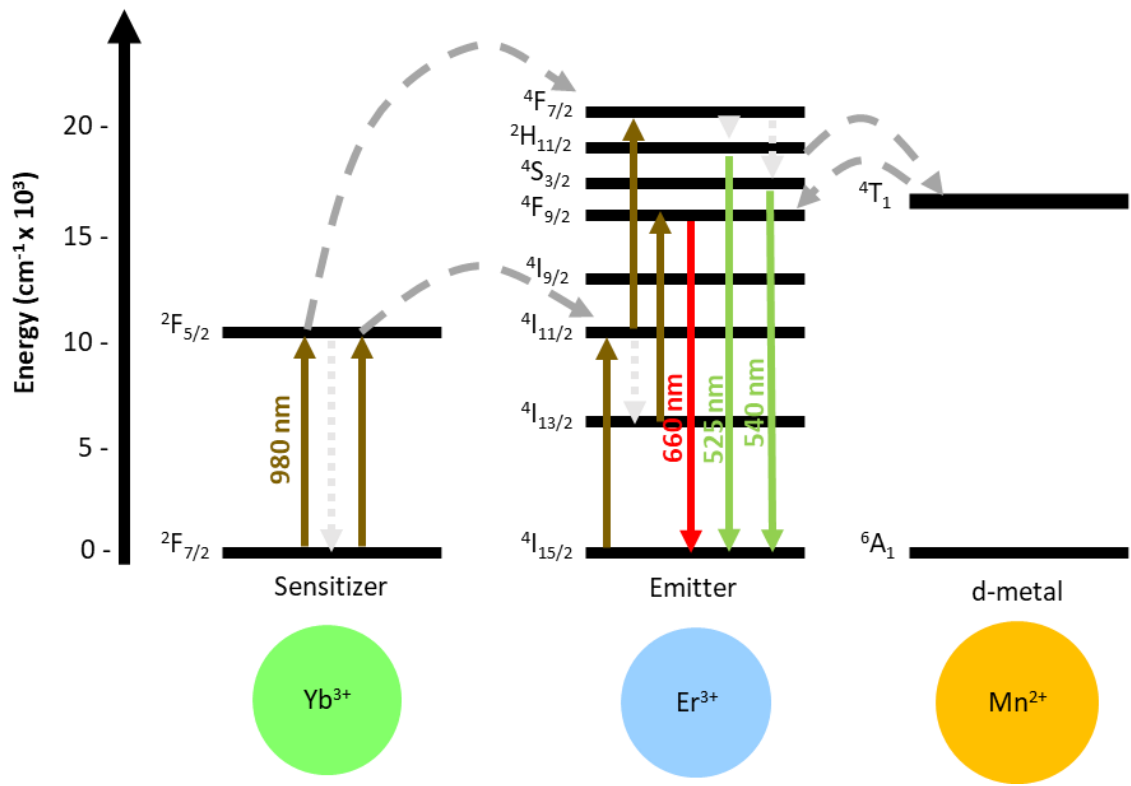

Figure S7. Energy level diagram with proposed energy transfer pathways leading from 980 nm excitation to upconverted Erbium ion emission. Solid lines indicate excitation and emission, dashed lines indicate interionic energy transfer, dotted lines indicate nonradiative energy loss. [Carnall et al. 1978]

Figure S8. Scanning electron micrographs of polymers (a) epoxy, (b) PDMS, (c) Mn-alginate, and (d) Ca-alginate.

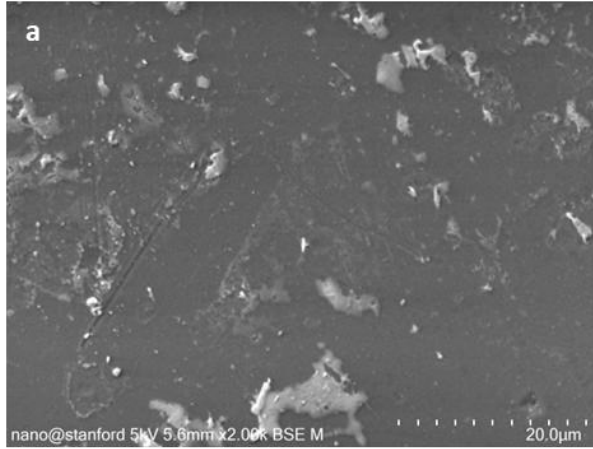
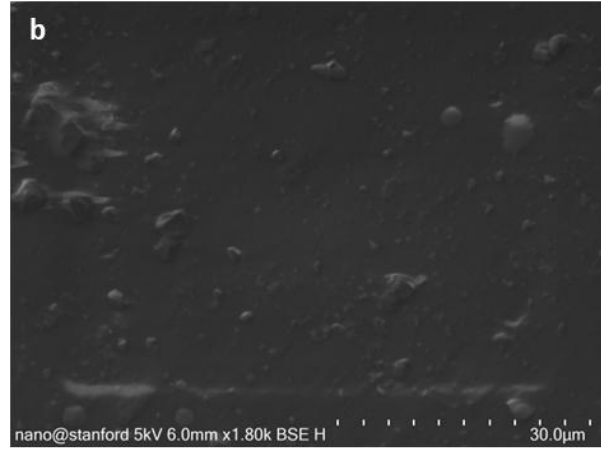
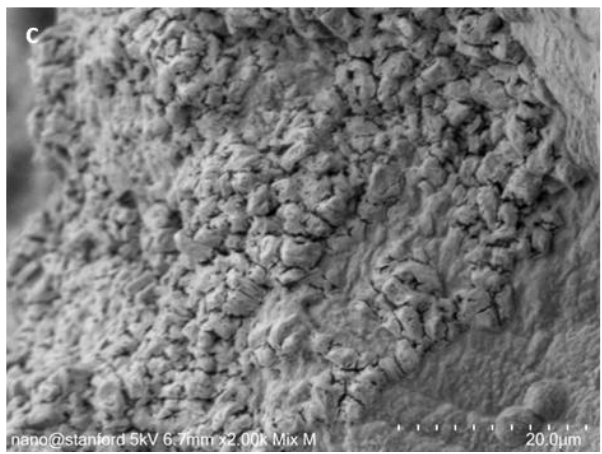
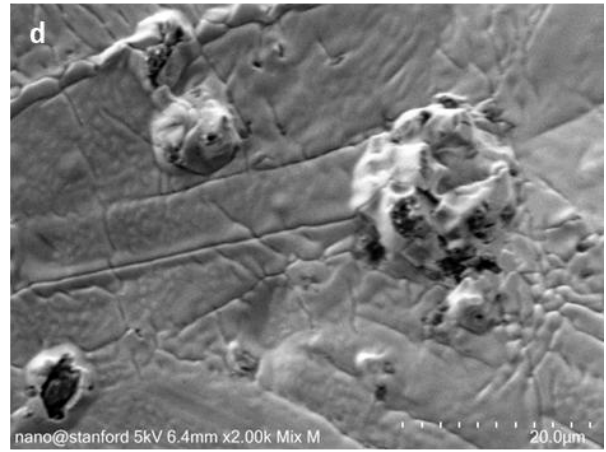

Figure S9. Mean red to green intensity of bulk epoxy-UCNP composite under calibration weights as a function of uniaxial compressive stress, measured on a wide field microscope.

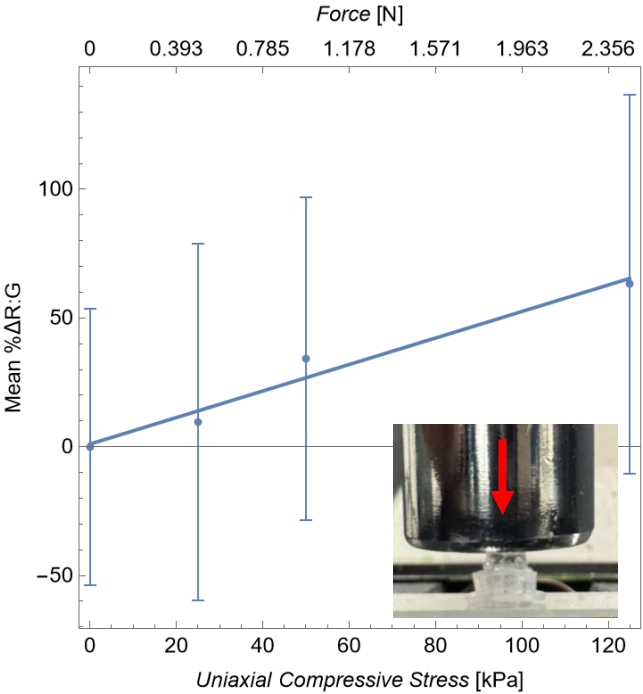

Figure S10. Force mapping on thin epoxy-UCNP coating inserted into the joint of a chicken wing bone on wide field microscope setup. (a) Schematic of imaging workflow, from sample illumination on wide field microscope to 3D force map. (b) 3D experimentally measured stress from red to green emission ratio maps of epoxy-UCNP coating between the joint of a chicken wing bone, with zero, two, and three loops of rubber band compression.

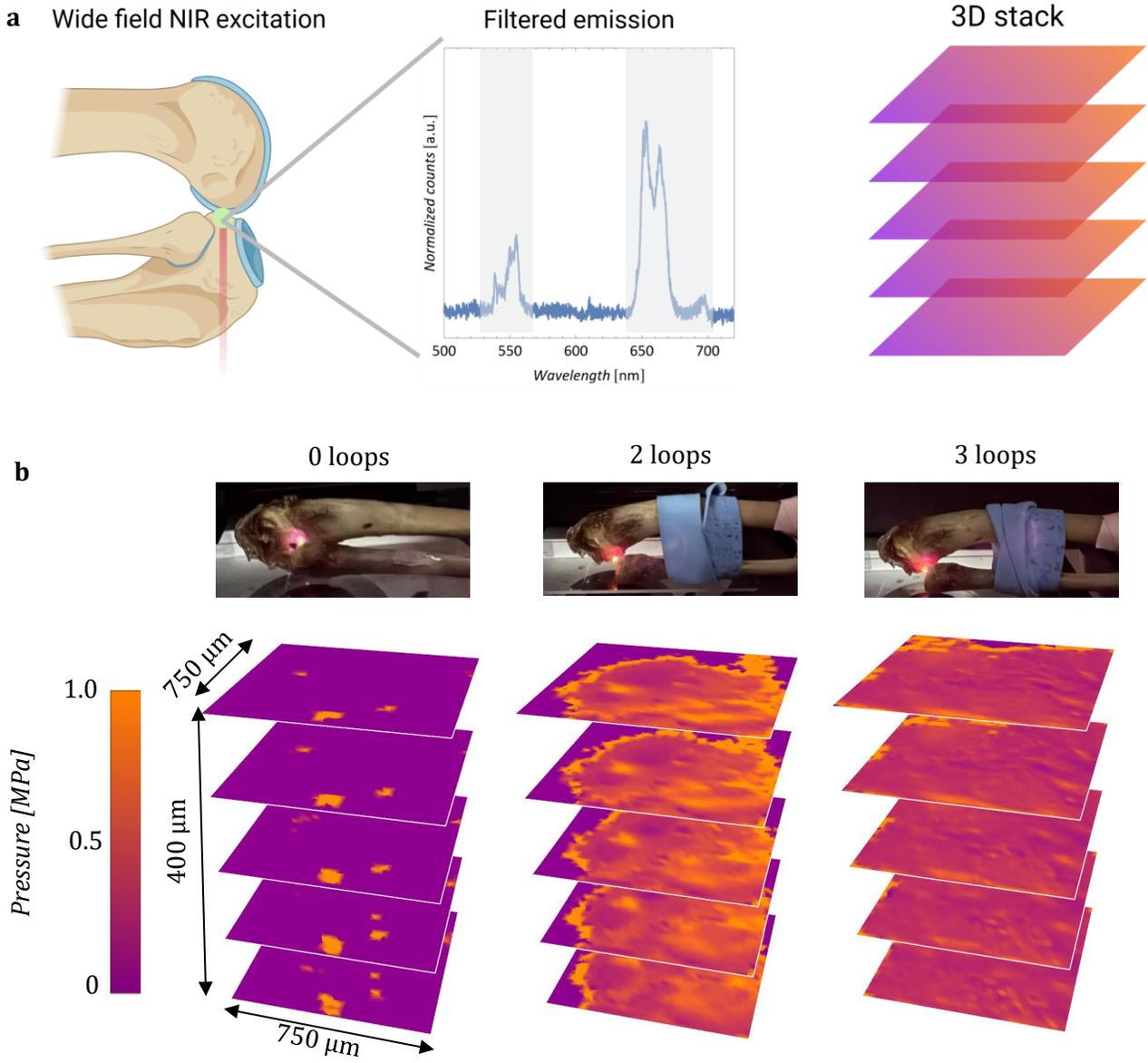